\documentclass{article}%
\usepackage{amsfonts}
\usepackage{amsmath}
\usepackage{geometry}
\usepackage{graphicx}
\usepackage{hyperref}
\usepackage{tabularx}
\usepackage{amssymb}%
\providecommand{\U}[1]{\protect\rule{.1in}{.1in}}
\begin{document}

\title{Exact Solution of Schrodinger Equation in (Anti-)deSitter Spaces for Hydrogen Atom}
\author{{Mokhtar FALEK}$^{1}$
\and {Noureddine BELGHAR}$^{2}$
\and {Mustafa MOUMNI}$^{1}$\\m.moumni@univ-biskra.dz\\$^{1}$Laboratory of Photonic Physics and Nano-Materials (LPPNNM)\\Department of Matter Sciences, University of Biskra, ALGERIA \\$^{2}$Laboratory of Energy Engineering and Materials (LGEM)\\Department of Mechanics, University of Biskra, ALGERIA}
\maketitle

\begin{abstract}
We write Schrödinger equation for the Coulomb potential in both deSitter and Anti-deSitter spaces using the Extended Uncertainty Principle formulation. We use the Nikiforov-Uvarov method to solve the equations. The energy eigenvalues for both systems are given in their exact forms and the corresponding radial wave functions are expressed in associated Jacobi polynomials for deSitter space, while those of Anti-deSitter space are given in terms of Romanovski polynomials. We have also studied the effect of the spatial deformation parameter on the bound states in the two cases.


\end{abstract}
\tableofcontents

\section{Introduction}

\label{Int}

The extension of the quantum field theory to curved space-time, which can be
considered as a first approximation of quantum gravity has attracted
considerable interest as there are strong motivations for absorption of
infinities lying in standard field theories. In such situation of curved
space-time, we deal with a structure perturbed by the gravitational field.
Such modifications can also be found in Snyder model where the measurements in
noncommutative quantum mechanics can be governed by a Generalized Uncertainty
Principle (GUP) \cite{Snyder}. This model admits a fundamental length scale
supposed to be of the order of the Planck length and this is equivalent to a
nonzero minimal uncertainty in the measurement of the position \cite{Kempf}%
\cite{Mendes}. Because there are many arguments showing that quantum gravity
implies also a minimal measurable length in the order of the Planck length, a
large amount of efforts have been devoted to extend the study of the quantum
mechanics to a curved space--time via the Extended Uncertainty Principle (EUP)
\cite{Chung}. A significant consequence deduced from this extension is that
the minimal length uncertainty in quantum gravity can be related also to a
modification of the standard Heisenberg algebra by adding small corrections to
the canonical commutation relations \cite{Mignemi}\cite{Ghosh}\cite{Nozari}.
This was motivated by Doubly Special Relativity (DSR) \cite{Amelino}, string
theory \cite{Capozziello}, non-commutative geometry \cite{Douglas} and also
black hole physics \cite{Scardigli}.

In the context of deformed quantum theory with EUP, there are only a few
available exact solutions. At the level of relativistic quantum mechanics the
list of the exactly solved problems is very restricted, e.g. the case of
one-dimensional Dirac and Klein-Gordon oscillators on anti-deSitter (AdS)
space was recently considered in \cite{Hamil1}, the three and two-dimensional
Dirac oscillator in the presence of minimal uncertainty in momentum was
studied in \cite{Hamil2} and the exact solution of (1+1)-dimensional bosonic
oscillator subject to the influence of an uniform electric field in AdS space
too \cite{Merad}. On the other hand, the non-relativistic case is also of
great interest and remains unexplored within this framework. Despite the fact that, in conventional field theory approach in static de Sitter and anti de Sitter space-time models, we cannot derive any nonrelativistic covariant Schr\"{o}dinger-like equation from covariant Klein-Fock-Gordon equation, we can use the EUP formulation to write the dS and AdS versions of the Schr\"{o}dinger equation. Indeed Hamil et al treat the exact solution of the D-dimensional Schrödinger equation for the free-particle and the harmonic-oscillator in AdS space \cite{Hamil4}. In \cite{Chung}, Chung study analytically the one dimensional box problem and the harmonic oscillator problem. Also in \cite{Ghosh}, Ghosh and Mignemi use perturbative methods to study both harmonic oscillator and Hydrogen atom.

Regarding the hydrogen atom and because of the physics that comes from
studying and understanding such system, there has been a growing interest in
the study of exact solutions of this kind of problem in the ordinary case
\cite{Reyes}\cite{Ran}\cite{Gordeyev}\cite{Tsutsui}\cite{Yepez} as well as
in\ the context of deformed quantum mechanics based on GUP and we cite here
the study of Schr\"{o}dinger equation for the Coulomb potential with minimal
length in one dimension \cite{Pedram}\cite{Nouicer}\cite{Fityo} and in three
dimensions \cite{Brau}\cite{Akhoury}\cite{Benczik}.

In this paper, we are looking for the analytical treatment of the hydrogen
atom when subject to gravitational effects governed by EUP because it has been
studied only perturbatively \cite{Ghosh}. For this purpose we solve the
non-relativistic Coulomb problem to get the exact form of the energy
eigenvalues and eigenfunctions. The paper is organized as follows: In Sec. II
\ref{Def}, we give a review dS and AdS models while In Sec. III \ref{N-U}, we
introduce Nikiforov-Uvarov (NU) method that we use to solve equation of our
system. We expose in Sec. IV \ref{Sch}, the explicit computations for the
hydrogen atom of the deformed Schr\"{o}dinger equation with EUP and in both dS
and AdS cases of the algebra. The energy eigenvalues are given in their exact
form and the corresponding radial wave functions are expressed in associated
Jacobi polynomials for dS space \ref{dS}\ and in terms of Romanovski
polynomials for the AdS space \ref{AdS}. In the end of this section, we
investigate numerically the spectroscopic implications of the EUP deformation.
Finally, the concluding remarks come in Sec. V \ref{cc}.

\section{Review on the Deformed Quantum Mechanics Relation}

\label{Def}

In three-dimensional space, the deformed Heisenberg algebra leading to EUP is
defined by the following commutation relations \cite{Mignemi1}\cite{Stetsko}
\begin{equation}
\left[  X_{i},X_{j}\right]  =0\text{ , }\left[  P_{i},P_{j}\right]
=i\hbar\tau\lambda\epsilon_{ijk}L_{k}\text{ , }\left[  X_{i},P_{j}\right]
=i\hbar\left(  \delta_{ij}-\tau\lambda X_{i}X_{j}\right)  \text{ with }%
\tau=-1,+1 \label{eqt1}%
\end{equation}
where $\lambda$ is the parameter of the deformation and it is very small
because, in the context of quantum gravity, this EUP parameter is determined
as a fundamental constant associated to the scale factor of the expanding
universe and it is proportional to the cosmological constant $\Gamma
=3\tau\lambda=3\tau/a^{2}$ where $a$ is the deSitter radius \cite{Bolen}.
$L_{k}$ is the component of the angular momentum expressed by:
\begin{equation}
L_{k}=\epsilon_{ijk}X_{i}P_{j} \label{eqt2}%
\end{equation}
and satisfying the usual algebra:%
\begin{equation}
\left[  L_{i},P_{j}\right]  =i\hbar\varepsilon_{ijk}P_{k}\text{ , }\left[
L_{i},X_{j}\right]  =i\hbar\varepsilon_{ijk}X_{k}\text{ , }\left[  L_{i}%
,L_{j}\right]  =i\hbar\varepsilon_{ijk}L_{k} \label{eqt3}%
\end{equation}

As in ordinary quantum mechanics, the commutation relation \ref{eqt1} gives
rise to a Heisenberg uncertainty relation:%
\begin{equation}
\Delta X_{i}\Delta P_{i}\geq\frac{\hbar}{2}\left(  1-\tau\lambda\left(  \Delta
X_{i}\right)  ^{2}\right)  \label{eqt4}%
\end{equation}
where we choose the states for which $\left\langle X_{i}\right\rangle =0$.

According to the value of $\tau$ we distinguish two kinds of subalgebra. For
$\tau=-1$, the deformed algebra is characterized by the presence of a nonzero
minimum uncertainty in momentum and it is called Anti-deSitter model. For
simplicity, we assume isotropic uncertainties $X_{i}=X$ and this allows us to
write the minimal uncertainty for the momentum in AdS model:%
\begin{equation}
\left(  \Delta P_{i}\right)  _{\min}=\hbar\sqrt{\tau\lambda} \label{eqt5}%
\end{equation}
For de Sitter model where $\tau=+1$, the relation \ref{eqt4} does not imply an
non-zero minimal value for momentum uncertainties.%

%TCIMACRO{\FRAME{ftbpFU}{3.659in}{2.3255in}{0pt}{\Qcb{Graphic of HUP and EUP in
%both dS and AdS Cases }}{\Qlb{fig1}}{fig1.eps}%
%{\special{ language "Scientific Word";  type "GRAPHIC";
%maintain-aspect-ratio TRUE;  display "USEDEF";  valid_file "F";
%width 3.659in;  height 2.3255in;  depth 0pt;  original-width 3.6115in;
%original-height 2.2857in;  cropleft "0";  croptop "1";  cropright "1";
%cropbottom "0";  filename 'fig1.eps';file-properties "XNPEU";}} }%
%BeginExpansion
\begin{figure}[ptb]%
\centering
\includegraphics[
height=2.3255in,
width=3.659in
]%
{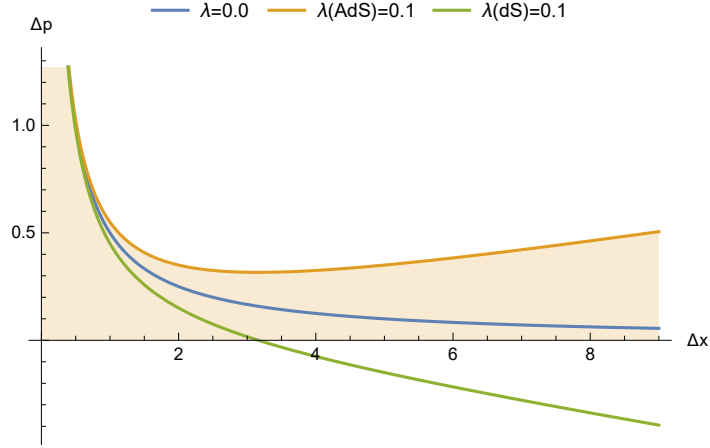}%
\caption{Graphic of HUP and EUP in both dS and AdS Cases }%
\label{fig1}%
\end{figure}
%EndExpansion

This is shown in figure1 \ref{fig1}, where the uncertainty relations are
plotted according to the modified relation found in \ref{eqt4}. The colored
region in \ref{fig1} is the forbidden area for position and momentum
measurements in AdS space.

The noncommutative operators $X_{i}$ and $P_{i}$ satisfy the modified algebra
\ref{eqt1} which gives rise to rescaled uncertainty relation \ref{eqt4} in
momentum space. In order to study the exact solutions of the deformed
Schr\"{o}dinger equation, we represent these operators as functions of the
operators $x_{i}$and $p_{i}$ that satisfy the ordinary canonical commutation
relations; This is done thanks to the following transformations:%
\begin{subequations}
\begin{align}
X_{i}  &  =\frac{x_{i}}{\sqrt{1+\tau\lambda r^{2}}}\label{eqt6a}\\
P_{i}  &  =-i\hbar\sqrt{1+\tau\lambda r^{2}}\partial_{x_{i}} \label{eqt6b}%
\end{align}
If $\tau=-1$, the variable $r$ varies in the domain $\left]  -1/\sqrt{\lambda
},1/\sqrt{\lambda}\right[  $.

\section{Nikiforov--Uvarov Method}

The Nikiforov-Uvarov (NU) method was developed basically on the hypergeometric
differential equation. The formulas used in NU method reduce the second order
differential equations to the hypergeometric type with an appropriate
coordinate transformation $s=s\left(  x\right)  $:
\end{subequations}
\begin{equation}
\psi^{\backprime\backprime}\left(  s\right)  +\frac{\tilde{\tau}\left(
s\right)  }{\sigma\left(  s\right)  }\psi^{\backprime}\left(  s\right)
+\frac{\tilde{\sigma}\left(  s\right)  }{\sigma^{2}\left(  s\right)  }%
\psi\left(  s\right)  =0 \label{eqt7}%
\end{equation}
where $\sigma\left(  s\right)  $ and $\widetilde{\sigma}\left(  s\right)  $
are polynomials of the second degree at most and the degree of the polynomial
$\widetilde{\tau}\left(  s\right)  $ is strictly less than 2 \cite{Egrifes}%
\cite{Nikiforov}. If we take the following factorization:%
\begin{equation}
\psi\left(  s\right)  =\phi\left(  s\right)  y\left(  s\right)  \label{eqt8}%
\end{equation}
\ref{eqt7} becomes \cite{Nikiforov}:%
\begin{equation}
\sigma\left(  s\right)  y^{\backprime\backprime}\left(  s\right)  +\tau\left(
s\right)  y^{\backprime}\left(  s\right)  +\Lambda y\left(  s\right)  =0
\label{eqt9}%
\end{equation}
where:%
\begin{equation}
\pi\left(  s\right)  =\sigma\left(  s\right)  \frac{d}{ds}\left(  \ln
\phi\left(  s\right)  \right)  \text{ \ and \ }\tau\left(  s\right)
=\tilde{\tau}\left(  s\right)  +2\pi\left(  s\right)  \label{eqt10}%
\end{equation}
$\Lambda$ is defined as:%
\begin{equation}
\Lambda_{n}+n\tau^{\backprime}+\frac{n\left(  n-1\right)  \sigma
^{\backprime\backprime}}{2}=0,\text{ \ \ }n=0,1,2,... \label{eqt11}%
\end{equation}
And the energy eigenvalues are calculated from the above equation. We first
have to determine $\pi\left(  s\right)  $ and $\Lambda$ by defining:%
\begin{equation}
k=\Lambda-\pi^{\backprime}\left(  s\right)  \label{eqt12}%
\end{equation}
Solving the quadratic equation for $\pi\left(  s\right)  $ with \ref{eqt12},
we get%
\begin{equation}
\pi\left(  s\right)  =\left(  \frac{\sigma^{\backprime}-\widetilde{\tau}}%
{2}\right)  \pm\sqrt{\left(  \frac{\sigma^{\backprime}-\tilde{\tau}}%
{2}\right)  ^{2}-\tilde{\sigma}+\sigma k} \label{eqt13}%
\end{equation}
Here, $\pi\left(  s\right)  $ is a polynomial of the parameter $s$ and the
prime denotes the first derivative.

One has to note that the determination of $k$ is the essential point in the
calculation of $\pi\left(  s\right)  $ and It is simply defined by stating
that the expression under the square root in \ref{eqt13} must be a square of a
polynomial; This gives us a general quadratic equation for $k$.

To determine the polynomial solutions $y_{n}\left(  s\right)  $, we use
\ref{eqt10} and the Rodrigues relation:%
\begin{equation}
y_{n}\left(  s\right)  =\frac{C_{n}}{\rho\left(  s\right)  }\frac{d^{n}%
}{ds^{n}}\left[  \sigma^{n}\left(  s\right)  \rho\left(  s\right)  \right]
\label{eqt14}%
\end{equation}
where $C_{n}$ is normalizable constant and the weight function $\rho\left(
s\right)  $ satisfies the following relation:%
\begin{equation}
\frac{d}{ds}\left[  \sigma\left(  s\right)  \rho\left(  s\right)  \right]
=\tau\left(  s\right)  \rho\left(  s\right)  \label{eqt15}%
\end{equation}
This last equation refers to the classical orthogonal polynomials that have
many important properties and especially orthogonality defined by:%
\begin{equation}
\int_{a}^{b}y_{n}\left(  s\right)  y_{m}\left(  s\right)  \rho\left(
s\right)  ds=0\text{ if }m\neq n \label{eqt16}%
\end{equation}

\section{Schr\"{o}dinger Equation for the Hydrogen Atom in (Anti-)deSitter
Space}

\label{Sch}

In this section, we study the effects of deformed space on the energy
eigenvalues and eigenfunctions of a hydrogen atom in the context of the
non-relativistic quantum mechanics. In the case of a three-dimensional
space,we consider the following stationary Schr\"{o}dinger equation with a
Coulomb-type interaction:
\begin{equation}
\left[  \frac{\mathbf{p}^{2}}{2m}-\frac{e^{2}}{r}\right]  \psi\left(
\mathbf{r}\right)  =E\psi\left(  \mathbf{r}\right)  \, \label{eqt17}%
\end{equation}

In order to include the effect of EUP on the above Schr\"{o}dinger equation,
we use the transformations \ref{eqt16a} and \ref{eqt6b} to obtain:%
\begin{equation}
\left[  -\frac{\hbar^{2}}{2m}\left[  \left(  1+\tau\lambda r^{2}\right)
\left(  \frac{\partial^{2}}{\partial r^{2}}+\frac{2}{r}\frac{\partial
}{\partial r}-\frac{L^{2}}{\hbar^{2}r^{2}}\right)  +\tau\lambda r\frac
{\partial}{\partial r}\right]  -\frac{e^{2}\sqrt{1+\tau\lambda r^{2}}}%
{r}\right]  \psi=E\psi\label{eqt18}%
\end{equation}
$\,$

In order to separate the variables, we write the solution as $\psi
(r,\theta)=r^{-1/2}R(r)Y_{l}^{m_{l}}(\theta,\varphi)$ and this enables us to
split the equation into two parts, one angular and the other radial (where
$\chi\equiv\sqrt{1+\tau\lambda r^{2}}$):%
\begin{equation}
L^{2}Y_{l}^{m_{l}}(\theta,\varphi)=\hbar^{2}l\left(  l+1\right)  Y_{l}^{m_{l}%
}(\theta,\varphi)\, \label{eqt19}%
\end{equation}%
\begin{equation}
\left[  \left(  \chi\frac{d}{dr}\right)  ^{2}+\frac{\chi^{2}}{r}\frac{d}%
{dr}-\frac{\left(  l\left(  l+1\right)  +\frac{1}{4}\right)  \chi^{2}}{r^{2}%
}+\frac{2me^{2}\chi}{\hbar^{2}r}\right]  R(r)=-\left(  \frac{2mE}{\hbar^{2}%
}+\frac{\tau\lambda}{2}\right)  R(r) \label{eqt20}%
\end{equation}
The angular equation of the system is just the usual one for spherical
harmonics, so we are interested in the resolution of the radial one. In order
to do this, we use the following transformations:%
\begin{equation}
s=\frac{\sqrt{1+\tau\lambda r^{2}}}{\sqrt{\lambda}r} \label{eqt21}%
\end{equation}
Then, the new form of \ref{eqt20} becomes:%
\begin{equation}
\left[  \left(  1-\tau s^{2}\right)  ^{2}\frac{d^{2}}{ds^{2}}-\tau s\left(
1-\tau s^{2}\right)  \frac{d}{ds}-\left(  l+\frac{1}{2}\right)  ^{2}s^{2}+\eta
s+\varepsilon\right]  R_{1,2}(s)=0\, \label{eqt22}%
\end{equation}
where%
\begin{equation}
\eta=\frac{2me^{2}}{\hbar^{2}\sqrt{\lambda}}\text{ and\ }\varepsilon
=\frac{2mE}{\lambda\hbar^{2}}+\frac{\tau}{2} \label{eqt23}%
\end{equation}

\subsection{Solutions for deSitter Space ($\tau=+1$)}

\label{dS}

Comparison between \ref{eqt22} and \ref{eqt7} allows us to use the NU method
where the expressions of the polynomials appearing in \ref{eqt7} are given by:%
\begin{equation}
\sigma\left(  s\right)  =\left(  1-s^{2}\right)  \text{ , }\tilde{\tau}\left(
s\right)  =-s\text{ \ and }\tilde{\sigma}\left(  s\right)  =-\left(
l+\frac{1}{2}\right)  ^{2}s^{2}+\eta s+\varepsilon\label{eqt24}%
\end{equation}
Substituting them into \ref{eqt13} we obtain:%
\begin{equation}
\pi\left(  s\right)  =\frac{-s}{2}\pm\sqrt{\left(  \frac{1}{4}+\left(
l+\frac{1}{2}\right)  ^{2}-k\right)  s^{2}-\eta s+k-\varepsilon} \label{eqt25}%
\end{equation}
Where the parameter $k$ is to be determined by the condition mentioned in the
section III \ref{N-U}. One then obtains the following possible solutions for
each $k$:%
\begin{equation}
\pi\left(  s\right)  =\left\{
\begin{array}
[c]{c}%
\pi_{1,2}=\left(  \frac{-1}{2}\pm\delta_{1}\right)  s\mp\frac{\eta}%
{2\delta_{1}}\text{\ for\ }k_{1}=\frac{1}{2}\left[  \varepsilon+\frac{1}%
{4}+\left(  l+\frac{1}{2}\right)  ^{2}+\sqrt{\Delta}\right] \\
\pi_{3,4}=\left(  \frac{-1}{2}\pm\delta_{2}\right)  s\mp\frac{\eta}%
{2\delta_{2}}\text{\ for\ }k_{2}=\frac{1}{2}\left[  \varepsilon+\frac{1}%
{4}+\left(  l+\frac{1}{2}\right)  ^{2}-\sqrt{\Delta}\right]
\end{array}
\right.  \label{eqt26}%
\end{equation}
with:%

\begin{equation}
\delta_{1,2}=\sqrt{\frac{1}{4}+\left(  l+\frac{1}{2}\right)  ^{2}-k_{1,2}%
}\text{ and }\Delta=\left(  \varepsilon-\frac{1}{4}-\left(  l+\frac{1}%
{2}\right)  ^{2}\right)  ^{2}-\eta^{2} \label{eqt27}%
\end{equation}
Here, we choose the proper value $\pi_{1}$, so that:%
\begin{equation}
\tau\left(  s\right)  =2\left(  \delta_{1}-1\right)  s-\frac{\eta}{\delta_{1}}
\label{eqt28}%
\end{equation}
From \ref{eqt11} we obtain:%
\begin{equation}
\Lambda=k_{1}-\frac{1}{2}+\delta_{1}=n_{r}\left(  n_{r}+1-2\delta_{1}\right)
\text{ , }n_{r}=0,1,2,... \label{eqt29}%
\end{equation}
Hence, the energy eigenvalues are found as:%
\begin{equation}
E_{n,l}=-\frac{me^{4}}{2\hbar^{2}n^{2}}-\frac{\lambda\hbar^{2}}{2m}\left(
n^{2}-l\left(  l+1\right)  -1\right)  \label{eqt30}%
\end{equation}
where $n=n_{r}+l+1$ is the principal quantum number.

We remark that the above expression of energies contains the usual Hydrogen
term and an additional correction term proportional to the deformation
parameter $\lambda$, so we recover the Bohr energies when the deformation
disappears. It should be noted here that the first term of the correction is
proportional to $n%
%TCIMACRO{\U{b2}}%
%BeginExpansion
{{}^2}%
%EndExpansion
$ and so it is equivalent to the energy of a non-relativistic quantum particle
moving in a square well potential; In our case, the boundaries of the well are
placed at $\pm\pi/2\sqrt{\lambda}$. The second term in the correction contains
the azimuthal quantum number $l$ and it removes the $\left(  2l+1\right)  $
degeneracy of the energy levels. We also notice that the correction
deformation affects all energy levels except the ground level ($n=1$) which
remains not affected by the deformation even for large values of $\lambda$.

Now let us find the corresponding eigenfunctions. Taking the expression of
$\pi_{1}\left(  s\right)  $ from \ref{eqt26}, the $\phi\left(  s\right)  $
part is defined from the relation \ref{eqt10} as below:
\begin{equation}
\phi\left(  s\right)  =\left(  1+s\right)  ^{\frac{1}{4}\left(  1-2\delta
_{1}-\frac{\eta}{\delta_{1}}\right)  }\left(  1-s\right)  ^{\frac{1}{4}\left(
1-2\delta_{1}+\frac{\eta}{\delta_{1}}\right)  } \label{eqt31}%
\end{equation}
and according to the form of $\sigma\left(  s\right)  $ \ref{eqt24}, the
$y\left(  s\right)  $ part is given by Rodrigues relation:
\begin{equation}
y_{n}\left(  s\right)  =\frac{C_{n}}{\rho\left(  s\right)  }\frac{d^{n}%
}{ds^{n}}\left[  \left(  1-s^{2}\right)  ^{n}\rho\left(  s\right)  \right]
\label{eqt32}%
\end{equation}
where $\rho\left(  s\right)  =\left(  1+s\right)  ^{\left(  -\delta_{1}%
-\frac{\eta}{2\delta_{1}}\right)  }\left(  1-s\right)  ^{-\left(  \delta
_{1}-\frac{\eta}{2\delta_{1}}\right)  }$. The expression \ref{eqt32} stands
for the Jacobi polynomials as:%
\begin{equation}
y_{n}\left(  s\right)  \equiv P_{n_{r}}^{\left(  -\delta_{1}-\frac{\eta
}{2\delta_{1}},-\delta_{1}+\frac{\eta}{2\delta_{1}}\right)  }\left(  s\right)
\label{eqt33}%
\end{equation}
Hence, $R_{dS}(s)$ can be written in the following form:%
\begin{equation}
R_{dS}(s)=C_{n}\left(  1-s\right)  ^{\frac{1}{4}\left(  1-2\delta_{1}%
+\frac{\eta}{\delta_{1}}\right)  }\left(  1+s\right)  ^{\frac{1}{4}\left(
1-2\delta_{1}-\frac{\eta}{\delta_{1}}\right)  }P_{n_{r}}^{\left(  -\delta
_{1}-\frac{\eta}{2\delta_{1}},-\delta+\frac{\eta}{2\delta_{1}}\right)
}\left(  s\right)  \label{eqt34}%
\end{equation}
In terms of the variables $r$, $\theta$ and $\varphi$, we can now write the
general form of the wave function $\Psi$ as follows:%
\begin{align}
\Psi_{n_{r}}\left(  r,\theta,\varphi\right)   &  =C_{n}\left(  1-\frac
{\sqrt{1+\lambda r^{2}}}{\sqrt{\lambda}r}\right)  ^{\frac{1}{4}\left(
1-2\delta_{1}+\frac{\eta}{\delta_{1}}\right)  }\left(  1+\frac{\sqrt{1+\lambda
r^{2}}}{\sqrt{\lambda}r}\right)  ^{\frac{1}{4}\left(  1-2\delta_{1}-\frac
{\eta}{\delta_{1}}\right)  }\times\nonumber\\
&  P_{n_{r}}^{\left(  -\delta_{1}-\frac{\eta}{2\delta_{1}},-\delta_{1}%
+\frac{\eta}{2\delta_{1}}\right)  }\left(  \frac{\sqrt{1+\lambda r^{2}}}%
{\sqrt{\lambda}r}\right)  Y_{l}^{m_{l}}(\theta,\varphi)\, \label{eqt35}%
\end{align}
where $C_{n}$ is a normalization constant.

\subsection{Solutions for Anti-deSitter Space ($\tau=-1$)}

\label{AdS}

By comparing \ref{eqt22} with \ref{eqt7}, we determine NU polynomials as
follows:%
\begin{equation}
\sigma\left(  s\right)  =\left(  1+s^{2}\right)  \text{ ,\ }\tilde{\tau
}\left(  s\right)  =s\text{ and }\tilde{\sigma}\left(  s\right)  =-\left(
l+\frac{1}{2}\right)  ^{2}s^{2}+\eta s+\varepsilon\label{eqt36}%
\end{equation}
Substituting them into \ref{eqt10}, we obtain:%
\begin{equation}
\pi\left(  s\right)  =\frac{s}{2}\pm\sqrt{\left(  k+\frac{1}{4}+\left(
l+\frac{1}{2}\right)  ^{2}\right)  s^{2}-\eta s+k-\varepsilon} \label{eqt37}%
\end{equation}
The constant $k$ is determined in the same way as in dS case. Therefore, we
get:%
\begin{equation}
\pi\left(  s\right)  =\left\{
\begin{array}
[c]{c}%
\pi_{1,2}=\left(  \frac{1}{2}\pm\delta_{1}^{\backprime}\right)  s\mp\frac
{\eta}{2\delta_{1}^{\backprime}}\text{\ for\ }k_{1}^{\backprime}=\frac{1}%
{2}\left[  \varepsilon-\frac{1}{4}-\left(  l+\frac{1}{2}\right)  ^{2}%
-\sqrt{\vartriangle^{\backprime}}\right] \\
\pi_{3,4}=\left(  \frac{1}{2}\pm\delta_{2}^{\backprime}\right)  s\mp\frac
{\eta}{2\delta_{2}^{\backprime}}\text{\ for\ }k_{2}^{\backprime}=\frac{1}%
{2}\left[  \varepsilon-\frac{1}{4}-\left(  l+\frac{1}{2}\right)  ^{2}%
+\sqrt{\vartriangle^{\backprime}}\right]
\end{array}
\right.  \label{eqt38}%
\end{equation}
where:%
\begin{equation}
\delta_{1,2}^{\backprime}=\sqrt{\frac{1}{4}+\left(  l+\frac{1}{2}\right)
^{2}+k_{1,2}^{\backprime}}\text{ and }\Delta^{\backprime}=\left(
\varepsilon+\frac{1}{4}+\left(  l+\frac{1}{2}\right)  ^{2}\right)  ^{2}%
+\eta^{2} \label{eqt39}%
\end{equation}
Here, we choose the proper value $\pi_{2}$, so that we have:%
\begin{equation}
\tau\left(  s\right)  =2\left(  1-\delta_{1}^{\backprime}\right)  s-\frac
{\eta}{\delta_{1}^{\backprime}} \label{eqt40}%
\end{equation}
From \ref{eqt11}, we calculate:%
\begin{equation}
\Lambda=k_{1}^{\backprime}+\frac{1}{2}-\sqrt{\frac{1}{4}+\left(  l+\frac{1}%
{2}\right)  ^{2}+k_{1}^{\backprime}}=-n_{r}\left(  n_{r}+1-2\sqrt{\frac{1}%
{4}+\left(  l+\frac{1}{2}\right)  ^{2}+k_{1}^{\backprime}}\right)
\label{eqt41}%
\end{equation}
Hence, the energy eigenvalues are found as:%
\begin{equation}
E_{n,l}=-\frac{me^{4}}{2\hbar^{2}n^{2}}+\frac{\lambda\hbar^{2}}{2m}\left(
n^{2}-l\left(  l+1\right)  -1\right)  \label{eqt42}%
\end{equation}

The same remarks made in the case of dS space apply here except that in this
case the correcting terms are inversely proportional to the deformation
parameter $\lambda$, so that the energies increases with increasing values
{}of $\lambda$ and the bound states in AdS space become less bounded than
those of the dS case for the same value of $\lambda$ \ref{eqt30}\ref{eqt42}.

In both cases, the spectral corrections due to EUP are qualitatively different
to those associated to GUP \cite{Brau}.

Now, to deduce the complete expression of the wave functions $\Psi_{n}\left(
x\right)  $, we use the expression \ref{eqt38} of $\pi_{2}\left(  s\right)  $
as follows:%
\begin{equation}
\phi\left(  s\right)  =\left(  1+s^{2}\right)  ^{\frac{1}{2}\left(  \frac
{1}{2}-\delta_{1}^{\backprime}\right)  }e^{\frac{-\eta}{2\delta_{1}%
^{\backprime}}\tan^{-1}\left(  s\right)  } \label{eqt43}%
\end{equation}
and using Rodrigues formula \ref{eqt14}, we find%
\begin{equation}
y_{n}\left(  s\right)  =\frac{C_{n}^{\backprime}}{\rho\left(  s\right)  }%
\frac{d^{n}}{ds^{n}}\left[  \left(  1+s^{2}\right)  ^{n}\rho\left(  s\right)
\right]  \label{eqt44}%
\end{equation}
where $\rho\left(  s\right)  =\left(  1+s^{2}\right)  ^{-\delta_{1}%
^{\backprime}}e^{\frac{\eta}{\delta_{1}^{\backprime}}\tan^{-1}\left(
s\right)  }$.

The relation \ref{eqt44} stands for the Romanovski polynomials \cite{Alvaro}
as:%
\begin{equation}
y_{n}\left(  s\right)  \equiv R_{n}^{\left(  -\delta_{1}^{\backprime}%
,\frac{-\eta}{\delta_{1}^{\backprime}}\right)  }\left(  s\right)  =\frac
{C_{n}^{\backprime}}{\left(  1+s^{2}\right)  ^{-\delta_{1}^{\backprime}%
}e^{\frac{-\eta}{\delta_{1}^{\backprime}}\tan^{-1}\left(  s\right)  }}%
\frac{d^{n}}{ds^{n}}\left[  \left(  1+s^{2}\right)  ^{n-\delta_{1}%
^{\backprime}}e^{\frac{-\eta}{\delta_{1}^{\backprime}}\tan^{-1}\left(
s\right)  }\right]  \label{eqt45}%
\end{equation}
Consequently, the expression of $R_{AdS}(s)$ is written as:%
\begin{equation}
R_{AdS}(s)=C_{n}\left(  1+s^{2}\right)  ^{\frac{1}{2}\left(  \frac{1}%
{2}-\delta_{1}\right)  }e^{\frac{-\eta}{2\delta_{1}}\tan^{-1}\left(  s\right)
}R_{n}^{\left(  -\delta_{1},\frac{-\eta}{\delta_{1}}\right)  }\left(
s\right)  \label{eqt46}%
\end{equation}
and the expression of the wave function $\Psi$ with the former variables $r$,
$\theta$ and $\varphi$ is given by:%
\begin{equation}
\Psi_{n}\left(  r,\theta,\varphi\right)  =C_{n}^{\backprime}\left(
\sqrt{\lambda}r\right)  ^{\left(  \delta_{1}^{\backprime}-\frac{1}{2}\right)
}e^{\frac{-\eta}{2\delta_{1}^{\backprime}}\tan^{-1}\left(  \frac
{\sqrt{1-\lambda r^{2}}}{\sqrt{\lambda}r}\right)  }R_{n}^{\left(  -\delta
_{1}^{\backprime},\frac{-\eta}{\delta_{1}^{\backprime}}\right)  }\left(
\frac{\sqrt{1-\lambda r^{2}}}{\sqrt{\lambda}r}\right)  Y_{l}^{m_{l}}%
(\theta,\varphi)\, \label{eqt47}%
\end{equation}
with $C_{n}^{\backprime}$ is a normalization constant.

In order to show the effects of the deformed Heisenberg algebra leading to EUP
on the bound-states of the\ Coulomb potential in non-relativistic quantum
mechanics systems, we plot, as an example, the energies levels of the
$s-$states $E_{n,0}$ versus the deformation parameters $\lambda$ for different
values of $n$ (we use the Hartree atomic units $m=\hbar=e=4\pi\varepsilon_{0}$
$=1$).%
%TCIMACRO{\FRAME{ftbpFU}{3.659in}{2.3255in}{0pt}{\Qcb{$E_{n,0}(\lambda)$ for
%$n=1,2$ \& $3$ in dS and AdS cases}}{\Qlb{fig2}}{fig2.eps}%
%{\special{ language "Scientific Word";  type "GRAPHIC";
%maintain-aspect-ratio TRUE;  display "USEDEF";  valid_file "F";
%width 3.659in;  height 2.3255in;  depth 0pt;  original-width 3.6115in;
%original-height 2.2857in;  cropleft "0";  croptop "1";  cropright "1";
%cropbottom "0";  filename 'fig2.eps';file-properties "XNPEU";}} }%
%BeginExpansion
\begin{figure}[ptb]%
\centering
\includegraphics[
height=2.3255in,
width=3.659in
]%
{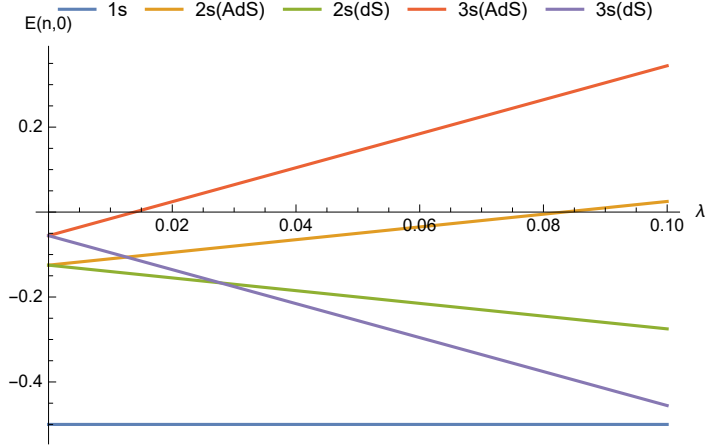}%
\caption{$E_{n,0}(\lambda)$ for $n=1,2$ \& $3$ in dS and AdS cases}%
\label{fig2}%
\end{figure}
%EndExpansion
According to the results shown in figure2 \ref{fig2} and to the expression of
the energies \ref{eqt42}, it is clear that the deformation increases the
energies in AdS case and thus decreases the binding energies of the states. We
thus arrive at a critical point where the value of the deformation parameter
cancels the bound state or $E_{n,l}=0$:%
\begin{equation}
\lambda_{c}(n,l)=\frac{1}{n^{2}\left(  n^{2}-l\left(  l+1\right)  -1\right)  }
\label{eqt48}%
\end{equation}
This critical values of the spatial deformation parameter can be interpreted
as a resonance point because the corresponding state of the atomic system
ionizes. We give in table1 \ref{tab1} some critical values $\lambda_{c}(n,l)$
corresponding to the first levels. Note from \ref{eqt30} that this is not the
case for dS space because the deformation increases the bonding of atomic
states and so no ionization effect occurs here.%
\begin{equation}%
\begin{tabular}
[c]{|c|c|c|c|c|c|}\hline
$\lambda_{c}(n,l)$ & $l=0$ & $l=1$ & $l=2$ & $l=3$ & $l=4$\\\hline
$n=2$ & $0.0833$ & $0.2500$ & $-$ & $-$ & $-$\\\hline
$n=3$ & $0.1389$ & $0.0185$ & $0.0556$ & $-$ & $-$\\\hline
$n=4$ & $0.0042$ & $0.0048$ & $0.0069$ & $0.0208$ & $-$\\\hline
$n=5$ & $0.0017$ & $0.0018$ & $0.0022$ & $0.0033$ & $0.0100$\\\hline
\end{tabular}
\ \tag{Table1}\label{tab1}%
\end{equation}

Figure2 \ref{fig2} and to the expression of the dS energies \ref{eqt30} show
that the deformation can reverse the order of energy levels since the
correction depends on the main quantum number. If we take the level $n=3$ as
an example, we see that it decreases faster than the $2^{nd}$ level and
therefore it becomes lower. Then it continues to decrease until it becomes
lower than the $1^{st}$ level, which will no longer be the fundamental one.
The value of $\lambda$ that causes this inversion between the upper levels and
the fundamental one is calculated from \ref{eqt30}:%
\begin{equation}
\lambda_{f}(n,l)=\frac{n^{2}-1}{n^{2}\left(  n^{2}-l\left(  l+1\right)
-1\right)  } \label{eqt49}%
\end{equation}

In table2 \ref{tab2}, we give some numerical values of $\lambda_{f}(n,l)$.%
\begin{equation}%
\begin{tabular}
[c]{|c|c|c|c|c|c|}\hline
$\lambda_{f}(n,l)$ & $l=0$ & $l=1$ & $l=2$ & $l=3$ & $l=4$\\\hline
$n=2$ & $0.250$ & $0.750$ & $-$ & $-$ & $-$\\\hline
$n=3$ & $0.111$ & $0.148$ & $0.444$ & $-$ & $-$\\\hline
$n=4$ & $0.063$ & $0.072$ & $0.104$ & $0.313$ & $-$\\\hline
$n=5$ & $0.040$ & $0.044$ & $0.053$ & $0.080$ & $0.240$\\\hline
\end{tabular}
\ \tag{Table2}\label{tab2}%
\end{equation}

Using \ref{eqt5}, we see that the main feature of the hydrogen spectrum in AdS
model \ref{eqt42} is the presence of an additional positive correction
proportional to the nonzero minimal uncertainty:%
\begin{equation}
E_{n,l}=-\frac{mk^{2}e^{4}}{2\hbar^{2}n^{2}}+\frac{\Delta P_{\min}^{2}}%
{2m}\left(  n^{2}-l\left(  l+1\right)  -1\right)  \label{eqt50}%
\end{equation}
One can use this relation to obtain an upper bound on the EUP deformed
parameter $\lambda\ $from spectroscopic considerations and we choose the
$2s-1s$\ transition line:%
\begin{equation}
\frac{E_{2s}-E_{1s}}{E_{1s}}=-\frac{3}{4}-\frac{3}{2}\frac{\hbar^{2}}%
{m^{2}k^{2}e^{4}}\Delta P_{\min}^{2} \label{eqt51}%
\end{equation}
Taking the experimental results for this transition in the hydrogen atom where
the precision is of the order of $\varepsilon\approx10^{-15}$ \cite{Matveev}
and if we attribute this error entirely to the EUP correction \ref{eqt50}, we
can write:%
\begin{equation}
\varepsilon=\frac{3}{2}\frac{\hbar^{2}}{m^{2}k^{2}e^{4}}\Delta P_{\min}^{2}
\label{eqt52}%
\end{equation}
Therefore, the upper bound of the minimal uncertainty in momentum is given by
$\Delta P_{\min}\sim10^{-32}Jsm^{-1}$; This value is much smaller than the one
obtained in \cite{Hamil2}.

\section{Conclusion}

\label{cc}

In this work, we have analytically studied the deformed Schr\"{o}dinger equation in three dimensions for a Hydrogen atom in deSitter and Anti-deSitter spaces by using the position representation of the Extended Uncertainty Principle formulation and the Nikiforov–Uvarov method. For both cases, we obtained
the exact eigen-energies and eigen-functions. The radial wave functions were
expressed as associated Jacobi polynomials for deSitter space and in terms of
Romanovski polynomials for Anti-deSitter space.

The deformed energy spectrum was written as the usual Coulomb term with an
additional correction term that removes the $l$ degeneracy of Bohr energies.
The main effect of the deformation parameter $\lambda$ is an increase of the
energies for AdS spaces and a decrease of these energies for dS spaces. It should be noted here that the two spectra are similar to those arising when considering the non-relativistic model for Hydrogen atom in space of constant negative and positive curvature: hyperbolic Lobachevsky and spherical Riemann models \cite{Redkov} (and the references therein).

In the AdS case, we showed that, due to the decrease of the binding energies
with increasing $\lambda$ values, we reach a critical point where the
corresponding state is no longer bound and thus becomes ionized or diffusive.
The critical values $\lambda_{c}(n,l)$ are inversely proportional to the
quantum numbers $n$ and $l$; so higher levels ionize one after the other as
$\lambda$ increases, until the stage where the atomic system contains only the
fundamental level. This is explained by the fact that the higher states are
more easily ionized even in the ordinary case.

On the other hand, in the case of dS space, all the energies levels are more
bounded proportionally to the values of the EUP parameter. Because bound
energies are increased according to $\lambda$ in this case, the deformation
can cause a reversal of the order of the levels where the energy of the higher
levels are diminished until becoming smaller than that of the fundamental
level. The corresponding values to this phenomenon $\lambda_{f}(n,l)$\ are
also inversely proportional to the quantum numbers $n$ and $l$.

These two effects of ionization and inversion are comparable to an extension
of the higher levels in the case AdS and a contraction of these same levels in
the case of dS.

Finally, in order to see the effect of the deformation on the physical
systems, we compared them with the experimental results of the
non-relativistic hydrogen atom and we have determined a satisfactory value of
the upper bound of minimal momentum uncertainty for AdS space.

\section*{Acknowledgment}

The authors would like to thank the referee for the remarks made; these have greatly improved the manuscript and thus contribute to a better understanding of the work.

\end{document}